\documentclass[11pt]{article}
\usepackage{graphicx}
\usepackage{amsmath}
\usepackage{amssymb}
\usepackage{amsxtra}

\newcommand{\lsim}{\mathrel{\mathop{\kern 0pt \rlap
  {\raise.2ex\hbox{$<$}}}
  \lower.9ex\hbox{\kern-.190em $\sim$}}}

\title{DAMA/LIBRA results and perspectives}
\author{R. Bernabei, P. Belli, A. Di Marco\\Dip. di Fisica, Univ. ``Tor Vergata'', I-00133 Rome, Italy\\
and INFN, sez. Roma ``Tor Vergata'', I-00133 Rome, Italy\\
F. Cappella, A. d'Angelo, A. Incicchitti\\Dip. di Fisica, Univ. di Roma ``La Sapienza'', I-00185 Rome, Italy\\
and INFN, sez. Roma, I-00185 Rome, Italy\\
V. Caracciolo, R. Cerulli\\Laboratori Nazionali del Gran Sasso, I.N.F.N., Assergi, Italy\\
C.J. Dai, H.L. He, X.H. Ma, X.D. Sheng, R.G. Wang\\IHEP, Chinese Academy, P.O. Box 918/3, Beijing 100039, China\\
F. Montecchia\\INFN, sez. Roma ``Tor Vergata'', I-00133 Rome, Italy\\
and Dip. di Ingegneria Civile e Ingegneria Informatica,\\
Univ. ``Tor Vergata'', I-00133 Rome, Italy\\
Z.P. Ye\\IHEP, Chinese Academy, P.O. Box 918/3, Beijing 100039, China\\
and University of Jing Gangshan, Jiangxi, China}

\begin{document}
\maketitle

\begin{abstract}
The DAMA/LIBRA experiment, running at the Gran Sasso National Laboratory
of the I.N.F.N. in Italy,
has a sensitive mass of about 250 kg highly radiopure
NaI(Tl). It is mainly devoted to the investigation
of Dark Matter (DM) particles in the Galactic halo by exploiting the model independent
DM annual modulation signature. The present DAMA/LIBRA experiment
and the former DAMA/NaI one (the first generation experiment having
an exposed mass of about 100 kg) have released so far
results corresponding to a total exposure of 1.17 ton $\times$ yr
over 13 annual cycles. They provide a model
independent evidence of the presence of DM particles in the galactic halo at 8.9 $\sigma$ C.L..
A short summary of the obtained results is presented and future perspectives of the
experiment mentioned.
\end{abstract}

\section{Introduction}

The DAMA project is an observatory for rare processes located deep 
underground at the Gran Sasso National Laboratory of the I.N.F.N.. 
It is based on the development and use of low
background scintillators. The main experimental set-ups are:
i) DAMA/NaI ($\simeq$ 100 kg of highly radiopure NaI(Tl)) that took data for 7 annual cycles 
and completed its data taking on July 2002 \cite{Nim98,allDM,Sist,RNC,ijmd,ijma,epj06,ijma07,chan,wimpele,ldm,allRare};
ii) DAMA/LXe, $\simeq$ 6.5 kg liquid Kr-free Xenon enriched either in $^{129}$Xe or
in $^{136}$Xe \cite{DAMALXe};
iii) DAMA/R\&D, a facility dedicated to test prototypes and to perform experiments 
developing and using various kinds of low background
crystal scintillators to investigate various rare processes \cite{DAMARD};
iv) DAMA/Ge, where sample measurements are carried out and where dedicated measurements 
on rare events are performed \cite{DAMAGE};
v) the second generation DAMA/LIBRA set-up, $\simeq$ 250 kg
highly radiopure NaI(Tl)) \cite{perflibra,modlibra,modlibra2,papep,cncn} mainly devoted to the investigation
of the presence of Dark Matter (DM) particles in the Galactic halo.
Profiting of the low background features of these set-ups, many rare
processes have been studied.

DAMA/LIBRA is the main apparatus, it is 
investigating the presence of DM particles in the galactic halo 
by exploiting the model independent DM annual modulation signature,
originally suggested in the mid 80's \cite{freese}.

In fact, as a consequence of its annual revolution around the Sun, 
which is moving in the Galaxy traveling with respect to the Local Standard 
of Rest towards the star Vega near
the constellation of Hercules, the Earth should be crossed by a larger 
flux of Dark Matter particles around $\sim$2 June 
(when the Earth orbital velocity is summed to the one of the
solar system with respect to the Galaxy) and by a smaller one 
around $\sim$2 December (when the two velocities are subtracted). 
Thus, this signature has a different origin and peculiarities
than the seasons on the Earth and than effects correlated with 
seasons (consider the expected value of the phase as well as the 
other requirements listed below). This DM annual modulation signature 
is very distinctive since the effect induced by DM particles must 
simultaneously satisfy all the following requirements:
(1) the rate must contain a component modulated according to a cosine
function; 
(2) with one year period; 
(3) with a phase that peaks 
roughly around $\sim$ 2nd June;
(4) this modulation
must be present only in a well-defined low energy
range, where DM particles can induce signals; (5) it must
be present only in those events where just a single detector,
among all the available ones in the used set-up, actually
``fires'' ({\it single-hit} events), since the probability that DM particles
experience multiple interactions is negligible; (6) the
modulation amplitude in the region of maximal sensitivity
has to be $\lsim$7$\%$ in case of usually adopted halo distributions,
but it may be significantly larger in case of some particular
scenarios such as e.g. those in refs. \cite{Wei01,Fre04}.

Only systematic effects or side reactions
able to simultaneously fulfill all the six requirements given 
above and to account for the whole observed modulation amplitude 
might mimic this DM signature; no one has been found or suggested 
by anyone over more than a decade.
Thus, no other effect investigated so far in the field of rare processes offers
a so stringent and unambiguous signature.

This offers an efficient model independent signature, able to test a large 
number of DM candidates, a large interval of
cross sections and of halo densities.
At present status of technology it is the only model independent signature available in 
direct Dark Matter investigation that can be effectively exploited.

It is worth noting that the corollary questions related to the exact nature of the DM
particle(s) (detected by means of the DM annual modulation signature)
and to the astrophysical, nuclear and particle Physics scenarios
require subsequent model dependent corollary analyses,
as those performed e.g. in refs. \cite{RNC,ijmd,ijma,epj06,ijma07,chan,wimpele,ldm}.
On the other hand, one should stress that it does not exist any approach
in direct and indirect DM
searches which can offer information on the nature of the candidate in a model independent way, 
that is without assuming any astrophysical, nuclear and particle Physics scenarios.

\section{DAMA/LIBRA results}

The DAMA/NaI set up and its performances are described in ref.\cite{Nim98,Sist,RNC,ijmd}, while
the DAMA/LIBRA set-up and its performances are described in ref. \cite{perflibra}.
The sensitive part of the DAMA/LIBRA set-up is made of 25
highly radiopure NaI(Tl) crystal scintillators placed in a 5-rows by 5-columns matrix;
each crystal is coupled to two low background photomultipliers working in 
coincidence at single photoelectron level. 
The detectors are placed inside a sealed copper box continuously flushed with HP nitrogen
and surrounded by a low background and massive shield made of  
Cu/Pb/Cd-foils/polyethylene/paraffin; moreover, about 1 m concrete (made from the Gran
Sasso rock material) almost fully surrounds (mostly outside
the barrack) this passive shield, acting as a further neutron
moderator. The installation has a 3-levels sealing system
which excludes the detectors from environmental air. The
whole installation is air-conditioned and the temperature is
continuously monitored and recorded. 
The detectors' responses range 
from 5.5 to 7.5 photoelectrons/keV. Energy calibrations with X-rays/$\gamma$ sources 
are regularly carried out down to few keV
in the same conditions as the production runs. In the data analysis a
software energy threshold of 2 keV is considered.

\begin{figure}[!h]
\vspace{-0.3cm}
\resizebox{\columnwidth}{!}{%
\includegraphics{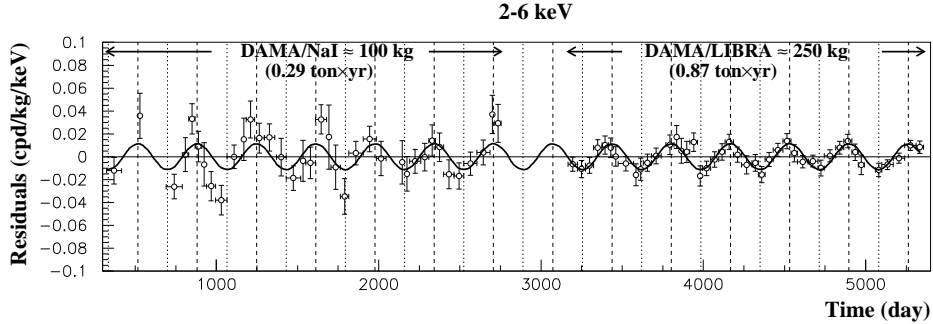}}
\vspace{-0.9cm}
\caption{Experimental model-independent residual rate of the {\it single-hit} scintillation events,
measured by DAMA/NaI over seven and by DAMA/LIBRA over six annual cycles in the (2 -- 6) keV energy interval
as a function of the time \protect\cite{RNC,ijmd,modlibra,modlibra2}. The zero of the time scale is January 1$^{st}$
of the first year of data taking.
The experimental points present the errors as vertical bars and the associated time bin width as horizontal bars.
The superimposed curve is $A \cos \omega(t-t_0)$
with period $T = \frac{2\pi}{\omega} =  1$ yr, phase $t_0 = 152.5$ day (June 2$^{nd}$) and
modulation amplitude, $A$, equal to the central value obtained by best fit over the whole data:
cumulative exposure is 1.17 ton $\times$ yr. The dashed vertical lines
correspond to the maximum expected for the DM signal (June 2$^{nd}$), while
the dotted vertical lines correspond to the minimum. See Refs.~\protect\cite{modlibra,modlibra2} and text.}
\label{fig1}
\end{figure}

The DAMA/LIBRA data released so far correspond to
six annual cycles for an exposure of 0.87 ton$\times$yr \cite{modlibra,modlibra2}.
Considering these data together with those previously collected by DAMA/NaI
over 7 annual cycles (0.29 ton$\times$yr), the total exposure collected
over 13 annual cycles is 1.17 ton$\times$yr; this
is orders of magnitude larger than the exposures typically collected in the field.
Several analyses on the model-independent DM annual
modulation signature have been performed (see Refs.~\cite{modlibra,modlibra2} and references therein);
here just few arguments are mentioned.
In particular, Fig. \ref{fig1} shows the time behaviour of the experimental
residual rates of the {\it single-hit}
events collected by DAMA/NaI and by DAMA/LIBRA in the (2--6) keV energy interval \cite{modlibra,modlibra2}.
The superimposed curve is the cosinusoidal function: $A \cos \omega(t-t_0)$
with a period $T = \frac{2\pi}{\omega} =  1$ yr, with a phase $t_0 = 152.5$ day (June 2$^{nd}$),
and modulation amplitude, $A$, obtained by best fit over the 13 annual cycles.
The hypothesis of absence of modulation in the data can be discarded \cite{modlibra,modlibra2} and,
when the period and the phase are released in the fit, values well compatible
with those expected for a DM particle induced effect are obtained \cite{modlibra2}; for example,
in the cumulative (2--6) keV energy interval:
$A = (0.0116 \pm 0.0013)$ cpd/kg/keV, $T = (0.999 \pm 0.002)$ yr and $t_0 = (146 \pm 7)$ day.
Summarizing, the analysis of the {\it single-hit} residual rate favours the presence of a
modulated cosine-like behaviour with proper features at 8.9 $\sigma$ C.L.\cite{modlibra2}.

\begin{figure}[!h]
\begin{center}
\vspace{-0.4cm}
\resizebox{\columnwidth}{!}{%
\includegraphics {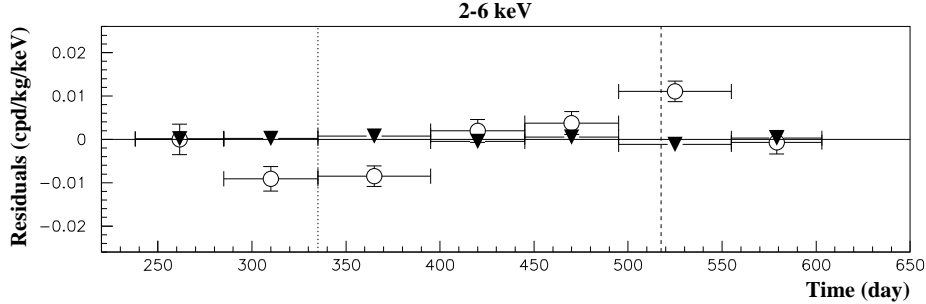}}
\vspace{-0.9cm}
\caption{Experimental residual rates over the six DAMA/LIBRA annual cycles for {\it single-hit} events
(open circles) (class of events to which DM events belong) and for {\it multiple-hit} events (filled triangles)
(class of events to which DM events do not belong).
They have been obtained by considering for each class of events the data as collected in a
single annual cycle
and by using in both cases the same identical hardware and the same identical software procedures.
The initial time of the figure is taken on August 7$^{th}$.
The experimental points present the errors as vertical bars and the associated time bin width as horizontal
bars. See text and Refs.~\protect\cite{modlibra,modlibra2}.
}
\label{fig_mul}
\vspace{-0.3cm}
\end{center}
\end{figure}

The same data of Fig.\ref{fig1} have also been investigated by a Fourier analysis, obtaining 
a clear peak corresponding to a period of 1 year \cite{modlibra2}; 
this analysis in other energy regions shows instead only aliasing peaks.
Moreover, while in the (2--6) keV {\it single-hit} residuals
a clear modulation is present, 
it is absent at energies just above \cite{modlibra2}. In particular, in order 
to verify absence of annual modulation in other energy regions and, thus,  
to also verify the absence of any significant background modulation, 
the energy distribution measured during the data taking periods
in energy regions not of interest for DM detection 
has also been investigated.
In fact, the background in the lowest energy region is
essentially due to ``Compton'' electrons, X-rays and/or Auger
electrons, muon induced events, etc., which are strictly correlated
with the events in the higher energy part of the spectrum;
thus, if a modulation detected 
in the lowest energy region would be due to
a modulation of the background (rather than to a signal),
an equal or larger modulation in the higher energy regions should be present.
The data analyses have allowed to exclude the presence of a background
modulation in the whole energy spectrum at a level much
lower than the effect found in the lowest energy region for the {\it single-hit} events
\cite{modlibra2}. 
A further relevant investigation has been done by applying the same hardware and software 
procedures,  used to acquire and to analyse the {\it single-hit} residual rate, to the 
{\it multiple-hits} events in which more than one detector ``fires''.
In fact, since the probability that a DM particle interacts in more than one detector 
is negligible, a DM signal can be present just in the {\it single-hit} residual rate.
Thus, this allows the study of the background behaviour in the same energy interval of the observed 
positive effect. The result of the analysis is reported in Fig. \ref{fig_mul} where 
it is shown the residual rate of the {\it single-hit} 
events measured over the six 
DAMA/LIBRA annual cycles, as collected in a single annual cycle, 
together with the residual rates of the {\it 
multiple-hits} events, in the same considered energy interval. 
A clear modulation is present in the {\it single-hit} events, while
the fitted modulation amplitudes for the {\it multiple-hits} 
residual rate are well compatible with zero \cite{modlibra2}.
Similar results were previously obtained also for the DAMA/NaI case \cite{ijmd}.
Thus, again evidence of annual modulation with proper features, as required by 
the DM annual modulation signature, is 
present in the {\it single-hit} residuals (events class to which the 
DM particle induced events belong), while it is absent in the {\it multiple-hits} residual rate (event class to 
which only background events belong).
Since the same identical hardware and the same identical software procedures have been used to analyse the 
two classes of events, the obtained result offers an additional strong support for the presence of a DM 
particle component in the galactic halo further excluding any side effect either from hardware or from software 
procedures or from background.

\begin{figure*}[!ht]
\begin{center}
%\vspace{-0.3cm}
\resizebox{0.75\columnwidth}{!}{%
\includegraphics {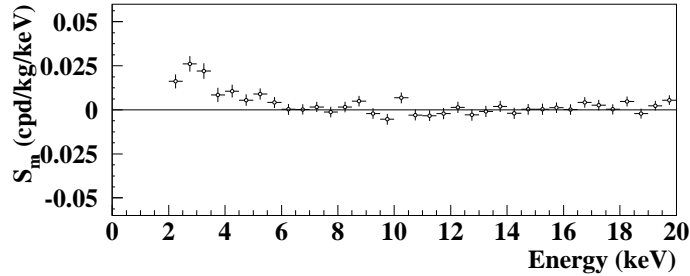}}
\vspace{-0.3cm}
\caption{Energy distribution of the modulation amplitudes $S_{m}$ for the
total cumulative exposure 1.17 ton$\times$yr. The energy bin is 0.5 keV.
A clear modulation is present in the lowest energy region,
while $S_{m}$ values compatible with zero are present just above. In fact, the $S_{m}$ values
in the (6--20) keV energy interval have random fluctuations around zero with
$\chi^2$ equal to 27.5 for 28 degrees of freedom.
See Refs.~\protect\cite{modlibra,modlibra2}.}
\label{fig_sm}
\vspace{-0.5cm}
\end{center}
\end{figure*}

The annual modulation present at low energy has also been analyzed 
by depicting the differential modulation amplitudes, 
$S_{m}$, as a function of the energy; the $S_{m}$ is the
modulation amplitude of the modulated part of the signal obtained
by maximum likelihood method over the data, considering $T=1$ yr and $t_0=152.5$ day.
The $S_{m}$ values are reported as function of the energy in Fig. \ref{fig_sm}.
It can be inferred that a positive signal is present in the (2--6) keV energy interval, while $S_{m}$
values compatible with zero are present just above; in particular, the $S_{m}$ values
in the (6--20) keV energy interval have random fluctuations around zero with
$\chi^2$ equal to 27.5 for 28 degrees of freedom.
It has been also verified that the measured modulation amplitudes are statistically well
distributed in all the crystals, in all the annual cycles and energy bins;
these and other discussions can be found in ref. \cite{modlibra2}.

It is also interesting the results of the analysis performed by releasing the 
assumption of a phase $t_0=152.5$ day in the procedure of maximum likelihood to 
evaluate the modulation amplitudes from the data of the
seven annual cycles of DAMA/NaI and the six annual cycles of DAMA/LIBRA. In this case
alternatively the signal has been written as:
$S_{0,k} + S_{m,k} \cos\omega(t-t_0) + Z_{m,k} \sin\omega(t-t_0) = 
 S_{0,k} + Y_{m,k} \cos\omega(t-t^*)$, where 
$S_{0,k}$ and $S_{m,k}$ are the constant part and the modulation amplitude
of the signal in $k$-th energy interval.
Obviously, for signals induced by DM particles one would expect: 
i) $Z_{m,k} \sim 0$ (because of the orthogonality between the cosine and the sine functions); 
ii) $S_{m,k} \simeq Y_{m,k}$; iii) $t^* \simeq t_0=152.5$ day. 
In fact, these conditions hold for most of the dark halo models; however, it is worth noting that 
slight differences in the phase could be expected in case of possible contributions
from non-thermalized DM components, such as e.g. the SagDEG stream \cite{epj06} 
and the caustics \cite{caus}.
The $2\sigma$ contours in the plane $(S_m , Z_m)$ 
for the (2--6) keV and (6--14) keV energy intervals and 
those in the plane $(Y_m , t^*)$ are reported in \cite{modlibra2}.
The best fit values for the (2--6) keV energy interval are ($1\sigma$ errors): 
$S_m= (0.0111 \pm 0.0013)$ cpd/kg/keV; 
$Z_m=-(0.0004 \pm 0.0014)$ cpd/kg/keV; 
$Y_m= (0.0111 \pm 0.0013)$ cpd/kg/keV; 
$t^*= (150.5  \pm 7.0)$ day;
while for the (6--14) keV energy interval are:
$S_m= -(-1.0001 \pm 0.0008)$ cpd/kg/keV;
$Z_m= (0.0002 \pm 0.0005)$ cpd/kg/keV;
$Y_m= -(0.0001 \pm 0.0008)$ cpd/kg/keV
and $t^*$ obviously not determined. 
These results confirm those achieved by other kinds of analyses.
In particular, a modulation amplitude is present in the lower energy intervals and the period
and the phase agree with those expected for DM induced signals.
For more detailed discussions see ref. \cite{modlibra2}

Both the data of DAMA/LIBRA and of DAMA/NaI
fulfil all the requirements of the DM annual modulation signature. 

Sometimes naive statements were put forwards as the fact that
in nature several phenomena may show some kind of periodicity.
It is worth noting that the point is whether they might
mimic the annual modulation signature in DAMA/LIBRA (and former DAMA/NaI), i.e. whether they
might be not only quantitatively able to account for the observed
modulation amplitude but also able to contemporaneously
satisfy all the requirements of the DM annual modulation signature; the same is also for side reactions.

Careful investigations
on absence of any significant systematics or side reaction able to
account for the measured modulation amplitude and to simultaneously satisfy
all the requirements of the signature
have been quantitatively carried out (see e.g. ref.
\cite{RNC,ijmd,modlibra,scineghe09,taupnoz,vulca010,muons12,answer}, refs therein). 
No systematics or side reactions able to mimic the signature (that is, able to
account for the measured modulation amplitude and simultaneously satisfy 
all the requirements of the signature) has been found or suggested 
by anyone over more than a decade. 

\begin{figure}[!ht]
\vspace{-.5cm}
\centering
\resizebox{0.5\columnwidth}{!}{%
\includegraphics{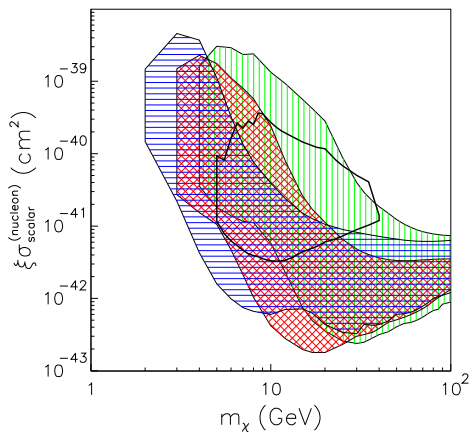} }
\vspace{-.5cm}
\caption{Regions in the nucleon cross section vs DM particle mass plane 
allowed by DAMA in three different instances for the Na and I quenching 
factors: i) without including the channeling effect [(green) 
vertically-hatched region], ii) by including the channeling effect 
[(blue) horizontally-hatched region)], and iii) without the channeling 
effect using the energy-dependent Na and I quenching factors \cite{bot11} [(red) 
cross-hatched region]. The velocity distributions and the same uncertainties 
as in Refs. \cite{RNC,ijmd} are considered here. The allowed region obtained for 
the CoGeNT experiment, including the same astrophysical models as in 
Refs. \cite{RNC,ijmd} and assuming for simplicity a fixed value for the Ge 
quenching factor and a Helm form factor with fixed parameters, is 
also reported and denoted by a (black) thick solid line. For details see Ref. \cite{bot11}.
}
\label{fig_mu}
\vspace{-0.1cm}
\end{figure}

The obtained model independent evidence
is compatible with a wide set of scenarios regarding the nature of the DM candidate
and related astrophysical, nuclear and particle Physics. For examples
some given scenarios and parameters are discussed e.g. in
Refs.~\cite{allDM,RNC,ijmd,ijma,epj06,ijma07,chan,wimpele,ldm} 
and in Appendix A of Ref.~\cite{modlibra}.
Further large literature is available on the topics; other possibilities are open.
Here we just recall the recent papers \cite{bot11,bot12} where the 
DAMA/NaI and DAMA/LIBRA results, which fulfill all the many peculiarities of 
the model independent DM annual modulation signature, are examined under the 
particular hypothesis of a light-mass DM candidate particle interacting with 
the detector nuclei by coherent elastic process. In particular, in Ref. \cite{bot11} allowed 
regions are given for DM candidates interacting by elastic scattering on nuclei 
including some of the existing uncertainties; comparison with theoretical 
expectations for neutralino candidate and with the recent possible positive 
hint by CoGeNT \cite{CoGeNT} is also discussed there (see Fig. 5), while comparison 
with possible positive hint by Cresst \cite{cresst} is discussed in Ref. \cite{bot12}. 

It is worth noting that no experiment exists, whose result can be directly compared in a 
model-independent way with those by DAMA/NaI and DAMA/LIBRA.

Some activities (e.g. \cite{cdms,edelweiss,xenon}) claim model-dependent exclusion under many 
largely arbitrary assumptions 
(see for example discussions in \cite{RNC,modlibra,ijmd,paperliq,collar});
often some critical points also exist in their experimental aspects 
(e.g. use of marginal exposures, determination of the energy threshold, of the energy resolution 
and of the energy scale in the few keV energy region of interest, 
multiple selection procedures, 
non-uniformity of the detectors response, absence of suitable periodical 
calibrations in the same running conditions and in the claimed low energy region, 
stabilities, tails/overlapping of the populations of the subtracted events and of the considered 
recoil-like ones, well known side processes mimicking recoil-like events, etc.); moreover,
the existing experimental and theoretical uncertainties are 
generally not considered in their presented model dependent result. 
Moreover, implications of the DAMA results are generally 
presented in incorrect/partial/unupdated way.

\section{Upgrades and perspectives}

A first upgrade of the DAMA/LIBRA set-up was performed in September 2008. 
One detector was recovered by replacing a broken PMT and a new optimization 
of some PMTs and HVs was done.
The transient digitizers were replaced with new ones, having better 
performances and a new DAQ with optical read-out was installed. 

A further and more important upgrade has been performed in the end of 2010 when all the 
PMTs have been replaced with new ones having higher quantum efficiency; details on the 
reached performances are reported in Ref. \cite{pmts}.
The purpose of the last upgrade of the running second generation DAMA/LIBRA 
set-up is: i) to increase the experimental sensitivity lowering the software
 energy threshold of the experiment; 2) to improve the investigation on the
 nature of the Dark Matter particle and related astrophysical, nuclear and 
particle physics arguments; 3) to investigate other signal features; 4) to 
improve the sensitivity in the investigation of rare processes other than 
Dark Matter as done by the former DAMA/NaI apparatus in the past \cite{allRare} and 
by itself so far \cite{papep,cncn}. This requires long and heavy full time dedicated 
work for reliable collection and analysis of very large exposures, 
as DAMA collaboration has always done.

Since January 2011 the DAMA/LIBRA experiment is again in data taking 
in the new configuration, named DAMA/LIBRA-phase 2.


\begin{thebibliography}{99}

%%%%%%%%%%%%%%%%% DAMA/NaI   %%%

\bibitem{Nim98}    R. Bernabei et al. Il Nuovo Cim. A {\bf 112}, 545 (1999).
\bibitem{allDM}    R. Bernabei et al. Phys. Lett. B {\bf 389}, 757 (1996); 
                   Phys. Lett. B {\bf 424}, 195 (1998); 
                   Phys. Lett. B {\bf 450}, 448 (1999); 
                   Phys. Rev.  D {\bf 61}, 023512 (2000);   
                   Phys. Lett. B {\bf 480}, 23 (2000);  
                   Phys. Lett. B {\bf 509}, 197 (2001); 
                   Eur. Phys. J. C {\bf 23}, 61 (2002); 
                   Phys. Rev. D {\bf 66}, 043503 (2002);
                   Int. J. Mod. Phys. A {\bf 21}, 1445 (2006);
                   Eur. Phys. J. C {\bf 47}, 263 (2006);
                   Int. J. Mod. Phys. A {\bf 22}, 3155 (2007);
                   Eur. Phys. J. C {\bf 53}, 205 (2008);
                   Phys. Rev. D {\bf 77}, 023506 (2008);
                   Mod. Phys. Lett. A {\bf 23}, 2125 (2008).

\bibitem{Sist}     R.~Bernabei et al., Eur. Phys. J. C {\bf 18}, 283 (2000).

%%%%%%%%%%%% DAMA/NaI DM
\bibitem{RNC}      R.~Bernabei et al. La Rivista del Nuovo Cimento {\bf26}, n.1, 1 (2003).
\bibitem{ijmd}     R.~Bernabei et al., Int. J. Mod. Phys. D {\bf 13}, 2127 (2004).

\bibitem{ijma}     R.~Bernabei et al., Int. J. Mod. Phys. A {\bf 21}, 1445 (2006).
\bibitem{epj06}    R.~Bernabei et al., Eur. Phys. J. C {\bf 47}, 263 (2006).
\bibitem{ijma07}   R.~Bernabei et al., Int. J. Mod. Phys. A {\bf 22}, 3155 (2007).
\bibitem{chan}     R.~Bernabei et al., Eur. Phys. J. C {\bf 53}, 205 (2008).
\bibitem{wimpele}  R.~Bernabei et al., Phys. Rev. D {\bf 77}, 023506 (2008).
\bibitem{ldm}      R.~Bernabei et al., Mod. Phys. Lett. A {\bf 23}, 2125 (2008).

%%%%%%%%%%%% DAMA/NaI rare processes

\bibitem{allRare}  See in the publication list in: http://people.roma2.infn.it/dama

%%%%%%%%%%%%DAMA/LXe%%%%%%%%%%%%%%%%%%%%%%%

\bibitem{DAMALXe} P. Belli et al., Astropart. Phys. {\bf 5}, 217 (1996);
                  Nuovo Cim. C {\bf 19}, 537 (1996);
                  Phys. Lett. B {\bf B387}, 222 (1996);
                  Phys. Lett. B {\bf 389}, 783 (err.) (1996);
                  R. Bernabei et al., Phys. Lett. B {\bf 436}, 379 (1998);
                  P. Belli et al., Phys. Lett. B {\bf 465}, 315 (1999);
                  Phys. Rev. D {\bf 61}, 117301(2000);
                  R. Bernabei et al., New J. of Phys. {\bf 2}, 15.1 (2000);
                  Phys. Lett. B {\bf 493}, 12 (2000);
                  Nucl. Instr. \& Meth A {\bf 482}, 728 (2000);    
                  Eur. Phys. J. direct C {\bf 11}, 1 (2001);
                  Phys. Lett. B {\bf 527}, 182 (2002);
                  Phys. Lett. B {\bf 546}, 23 (2002);
                  {\it Beyond the Desert 2003} (Berlin: Springer)  p. 365 (2003);
                  Eur. Phys. J. A {\bf 27}, s01 35 (2006).
                  
%%%%%%%%%%%%%%%%%%%%%%%%DAMA/R&D%%%%%%%%%

\bibitem{DAMARD}  R. Bernabei et al., Astropart. Phys. {\bf 7}, 73 (1997); 
                  Nuovo Cim. A {\bf 110}, 189 (1997); 
                  P. Belli et al., Astropart. Phys. {\bf 10}, 115 (1999);
                  Nucl. Phys. B {\bf 563}, 97 (1999);
                  R. Bernabei et al., Nucl. Phys. A {\bf 705}, 29 (2002);
                  P. Belli et al., Nucl. Instr. \& Meth A {\bf 498}, 352 (2003); 
                  R. Cerulli et al., Nucl. Instr. \& Meth A {\bf 525}, 535 (2004);
                  R. Bernabei et al., Nucl. Instr. \& Meth A {\bf 555}, 270 (2005);
                  Ukr. J. Phys. {\bf 51}, 1037 (2006);
                  P. Belli et al., Nucl. Phys. A {\bf 789}, 15 (2007);
                  Phys. Rev. C {\bf 76}, 064603 (2007);
                  Phys. Lett. B{\bf 658}, 193 (2008);
                  Eur. Phys. J. A {\bf 36}, 167 (2008); 
                  Nucl. Phys. A {\bf 826}, 256 (2009);
                  Nucl. Instr. \& Meth A {\bf 615}, 301 (2010); 
                  Nucl. Instr. \& Meth A {\bf 626-627}, 31 (2011); 
                  J. Phys. G: Nucl. Part. Phys. {\bf 38}, 015103 (2011).

%%%%%%%%%%%%%%%%%%%%%%%%DAMA/GE%%%%%%%

\bibitem{DAMAGE}  P. Belli et al., Nucl. Instr. \& Meth. A {\bf 572}, 734 (2007); 
                  Nucl. Phys. A {\bf 806}, 388 (2008); 
                  Nucl. Phys. A {\bf 824}, 101 (2009);
                  {\it Proceed. of the Int. Conf. NPAE 2008}
                  (ed. INR-Kiev, Kiev), p. 473 (2009);
                  Eur. Phys. J. A {\bf 42}, 171 (2009); 
                  Nucl. Phys.  A {\bf 846}, 143 (2010);
                  Nucl. Phys.  A {\bf 859}, 126 (2011);
                  Phys. Rev.   C {\bf 83}, 034603 (2011);              
                  Eur. Phys. J. A {\bf 47}, 91 (2011).
                  
%%%%%%%%%%%%%%%%%%%%%%%%DAMA/LIBRA%%%%%%%

\bibitem{perflibra} R. Bernabei et al., Nucl. Instr. \& Meth. A {\bf 592}, 297 (2008). 
\bibitem{modlibra}  R. Bernabei et al., Eur. Phys. J. C {\bf 56}, 333 (2008).
\bibitem{modlibra2} R. Bernabei et al., Eur. Phys. J. C {\bf 67}, 39 (2010).
\bibitem{papep}     R. Bernabei et al., Eur. Phys. J. C {\bf 62}, 327--332 (2009). 
\bibitem{cncn}      R. Bernabei et al., Eur. Phys. J. C {\bf 72}, 1920 (2012). 

\bibitem{freese}   A.K. Drukier, K. Freese, and D.N. Spergel, Phys. Rev. D {\bf 33}, 3495 (1986);
                   K. Freese, J. A. Frieman and A. Gould, Phys. Rev. D {\bf 37}, 3388 (1988).

\bibitem{Wei01}    D.~Smith and N.~Weiner, Phys. Rev. D {\bf 64}, 043502 (2001);
                   D.~Tucker-Smith and N.~Weiner, Phys. Rev. D {\bf 72}, 063509 (2005).

\bibitem{Fre04}    K. Freese et al., Phys. Rev. D {\bf 71}, 043516 (2005); 
                   Phys. Rev. Lett. {\bf 92}, 111301 (2004).

\bibitem{caus}     F.~S. Ling, P.~Sikivie and S.~Wick, Phys. Rev. D {\bf 70}, 123503--19 (2004).
\bibitem{scineghe09} R.~Bernabei et al., {\it AIP Conf. Proceed.}  {\bf 1223}, 50 (2010) [arXiv:0912.0660].
\bibitem{taupnoz}  R.~Bernabei et al., J. Phys.: Conf. Ser. {\bf 203}, 012040 (2010)
                   [arXiv:0912.4200];
                   http://taup2009.lngs.infn.it/slides/jul3/nozzoli. pdf, talk given by F. Nozzoli.
\bibitem{vulca010} R.~Bernabei et al., arXiv:1007.0595
                   to appear on {\it Proceed. of the Int. Conf. Frontier
                   Objects in Astrophysics and Particle Physics},
                   May 2010, Vulcano, Italy.

\bibitem{muons12}  R. Bernabei et al., Eur. Phys. J. C {\bf 72}, 2064 (2012). 
\bibitem{answer}   R. Bernabei et al., arXiv:1210.5501

\bibitem{bot11}    P. Belli et al., Phys. Rev. D {\bf 84}, 055014 (2011).

\bibitem{bot12}    A.~Bottino et al., Phys. Rev. D {\bf 85}, 095013 (2012).

\bibitem{CoGeNT}   C.E. Aalseth et al., arXiv:1002.4703; arXiv:1106.0650
\bibitem{cresst}   G. Angloher et al., arXiv:1109.0702

\bibitem{cdms}     Z. Ahmed et al., Science {\bf 327}, 1619 (2010).
\bibitem{edelweiss} E. Armengaud et al., Phys. Lett. B {\bf 702}, 329 (2011).
\bibitem{xenon}    E. Aprile et al., Phys. Rev. Lett. {\bf 105}, 131302 (2010).

\bibitem{paperliq} R. Bernabei et al., {\it Liquid Noble gases for Dark Matter searches: a synoptic survey},
                   Exorma Ed., Roma, ISBN 978-88-95688-12-1, pp. 1--53 (arXiv:0806.0011v2) (2009).
\bibitem{collar}   J.I. Collar and D.N. McKinsey, arXiv:1005.0838; arXiv:1005.3723;
                   J.I. Collar, arXiv:1006.2031; arXiv:1010.5187; arXiv:1103.3481; arXiv:1106.0653;
                   arXiv:1106.3559
\bibitem{pmts}     R. Bernabei et al., J. of Inst. {\bf 7}, P03009 (2012). 

\end{thebibliography}
\end{document}